%% file: arxiv.tex
\pdfoutput=1
\documentclass[sigconf]{acmart}
\usepackage{multirow}
\usepackage{footnote}
\usepackage{graphicx}
\usepackage{booktabs,caption}
\usepackage[flushleft]{threeparttable}
\usepackage{algorithm}
\usepackage{algorithmicx}
\usepackage{algpseudocode}
\usepackage{amsmath}
\usepackage{algpseudocode}
\usepackage{mathrsfs}
\usepackage{pifont}
\usepackage{url}
\usepackage{makecell}
\usepackage{subcaption}
\usepackage{enumitem}
\usepackage[usenames,dvipsnames]{colortbl}\usepackage{array}

\makeatletter
\newcommand{\thickhline}{%
    \noalign {\ifnum 0=`}\fi \hrule height 1pt
    \futurelet \reserved@a \@xhline
}
\definecolor{mycyan}{cmyk}{.8, 0, 0, 0}
\definecolor{gray-0}{rgb}{ .94, 1,1}
\definecolor{gray-1}{rgb}{ .90, 1,1}
\definecolor{gray-2}{rgb}{ .84, 1,1}

\begin{document}

\title{DeLoad: Demand-Driven Short-Video Preloading with Scalable Watch-Time Estimation}

\author{Tong Liu}
\affiliation{%
  \institution{Bytedance, China}
  \city{Beijing}
  \country{China}}
\email{tongliu@bytedance.com}

\author{Zhiwei Fan}
\affiliation{%
  \institution{Bytedance, China}
  \city{Beijing}
  \country{China}}
\email{fanzhiwei.rice@bytedance.com}

\author{Guanyan Peng}
\affiliation{%
  \institution{Bytedance, China}
  \city{Beijing}
  \country{China}}
\email{pengguanyan.shaw@bytedance.com}

\author{Haodan Zhang}
\affiliation{%
  \institution{Bytedance, China}
  \city{Beijing}
  \country{China}}
\email{zhanghaodan.2024@bytedance.com}

\author{Yucheng Zhang}
\affiliation{%
  \institution{Beijing University of Posts and Telecommunications, BUPT}
  \city{Beijing}
  \country{China}}
\authornote{Zhang's work is done during internship at ByteDance.}
\email{zhangyucheng@bupt.edu.cn}

\author{Zhen Wang}
\authornote{Corresponding author. Email: wangzhen3560@bytedance.com}
\affiliation{%
  \institution{Bytedance, China}
  \city{Beijing}
  \country{China}}
\email{wangzhen3560@bytedance.com}

\author{Pengjin Xie}
\affiliation{%
  \institution{Beijing University of Posts and Telecommunications, BUPT}
  \city{Beijing}
  \country{China}}
\email{xiepengjin@bupt.edu.cn}

\author{Liang Liu}
\affiliation{%
  \institution{Beijing University of Posts and Telecommunications, BUPT}
  \city{Beijing}
  \country{China}}
\email{liangliu@bupt.edu.cn}

\renewcommand{\shortauthors}{Tong Liu et al.}

\begin{abstract}
Short video streaming has become a dominant paradigm in digital media, characterized by rapid swiping interactions and diverse media content. A key technical challenge is designing an effective preloading strategy that dynamically selects and prioritizes download tasks from an evolving playlist, balancing \textit{Quality of Experience (QoE)} and \textit{bandwidth efficiency} under practical commercial constraints. However, real-world analysis reveals critical limitations of existing approaches: (1) insufficient adaptation of download task sizes to dynamic conditions, and (2) watch-time prediction models that are difficult to deploy reliably at scale. In this paper, we propose \textbf{DeLoad}, a novel preloading framework that addresses these issues by introducing dynamic task sizing and a practical, multi-dimensional watch-time estimation method. Additionally, a Deep Reinforcement Learning (DRL)-enhanced agent is trained to optimize the download range decisions adaptively. Extensive evaluations conducted on an offline testing platform, leveraging massive real-world network data, demonstrate that DeLoad achieves significant improvements in QoE metrics (34.4\%-87.4\% gain). Furthermore, after deployment on a large-scale commercial short-video platform, DeLoad has increased overall user watch-time by 0.9\text{\textperthousand} while simultaneously reducing rebuffering events and 3.76\% bandwidth consumption. 
\end{abstract}

\maketitle

\input{introduction}
\input{background}

\input{method}
\input{evaluation}

\input{relatedwork}

\input{conclusion}
\bibliographystyle{ACM-Reference-Format}
\balance
\bibliography{arxiv}
\end{document}

%% file: introduction.tex
\section{Introduction}\label{introduction}

In recent years, short-video platforms have emerged as the predominant paradigm in mobile streaming ecosystems~\cite{SensorTower2024}. Distinguished from traditional long-form VoD systems, these platforms leverage instantaneous swipe interactions to achieve markedly higher user engagement frequencies and enhanced immersive experiences~\cite{violot2024shorts, exp}. During playback, dynamically generated playlists enable users to seamlessly transition between videos based on personalized preferences, thereby introducing novel challenges for quality-of-experience (QoE) optimization. Unlike conventional single-video optimization frameworks, short-video downloading and preloading\footnote{In this paper, we regard \textbf{downloading} for current video and \textbf{preloading} for other videos as the equivalent concepts for the integrated short-video playlists} algorithms must address multi-dimensional QoE metrics-including playback smoothness, bitrate consistency, and cross-video switching latency—across the entire playlist hierarchy. This paradigm shift necessitates a fundamental rethinking of traditional preloading optimization methodologies.  

From an operational standpoint, bandwidth efficiency constitutes a critical economic bottleneck for commercial platforms. Inefficient preloading operations will raise extreme economic waste. Such inefficiencies not only escalate operational costs but also induce systemic trade-offs: excessive bandwidth allocation to low-priority content can degrade buffer stability for subsequent high-engagement videos, empirically increasing rebuffering probabilities under constrained network conditions. Consequently, optimizing bandwidth utilization emerges as a dual-objective problem that directly correlates with both user retention and profitability metrics.

At the core of short-video streaming systems lies the adaptive preloading algorithm, which operates in a high-dimensional decision space encompassing video prioritization, task sizes, and bitrate adaptation. This algorithm’s efficacy directly governs the cache situation (a primary determinant of QoE) and bandwidth consumption patterns. An optimally designed preloading strategy can elevate user watch time for the commercial platforms.

\begin{figure}[t]
  \centering
  \includegraphics[width=1\linewidth]{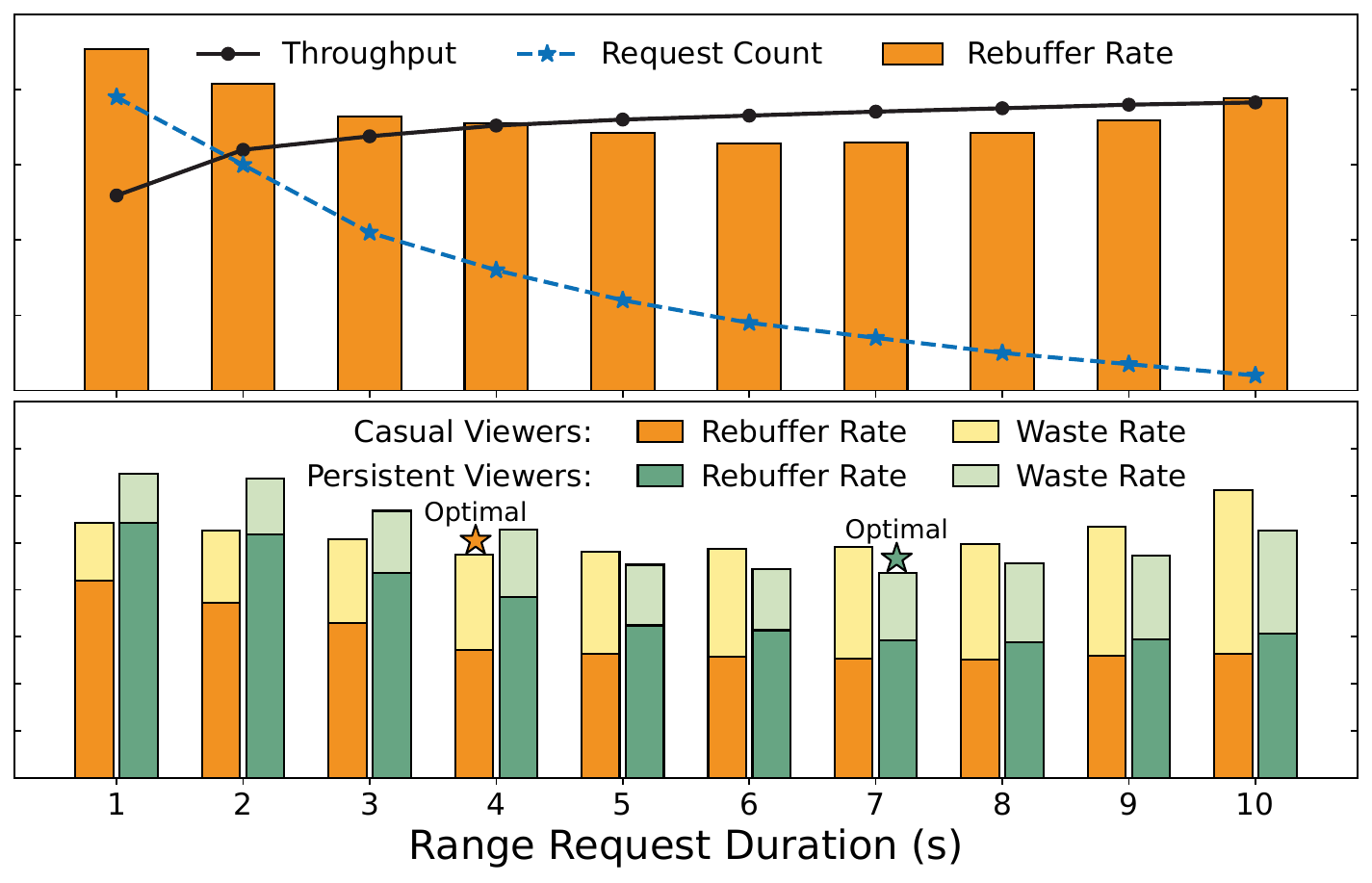}
  \caption{
    Impact analysis under different fixed range request durations. 
    \textit{Top}: Overall impact. 
    \textit{Bottom}: Comparative analysis between casual and persistent viewers. 
  }
  \label{fig:range-request-motivation}
\end{figure}

Current algorithms~\cite{qian2022dam, chao2022pdas, dashlet} predominantly follow rigid playlist-ordering strategies with fixed chunk sizes or \textit{duration lengths}, named as \textbf{range} for better representation lately, neglecting fine-grained network-aware optimization. This limitation becomes pronounced in mobile environments characterized by volatile bandwidth and heterogeneous content characteristics (e.g., variable video lengths and user watch durations). An example is shown in Fig.\ref{fig:range-request-motivation} (with details in Section~\ref{background}), the optimal range size varies across different user groups.
Furthermore, other approaches~\cite{Digitaltwin, gamora, dashlet, PACS, joint} rely heavily on user behavior predictions (e.g., swipe timing or final watch duration prediction) yet struggle with real-world uncertainties: content diversity, user heterogeneity, and inherent unpredictability among different engagement patterns. These factors render conventional static prediction models insufficient for robust downloading decisions. Consequently, existing algorithms exhibit clear limitations regarding further enhancement of QoE and reduction of data waste.

To address these challenges, we propose \textbf{DeLoad}, a novel preloading framework that achieves robust and practical improvements in both QoE and cost optimization. DeLoad consists of three main components:
\begin{itemize}[leftmargin=0pt, itemindent=1em]

\item \textbf{Robust Watch-Time Modeling and Prediction:} We design a novel watch-time modeling and prediction method based on the Weibull distribution \cite{weibull}. We further develop a multi-dimensional prediction approach that enables the estimation of a user's watch time for previously unseen videos, even under sparse data conditions.

\item \textbf{Demand-Driven Video Selection}: For short video scenarios, we define a probability-based metric called \textbf{Demand}, and develop a mathematical model for the design of the video selection algorithm. Demand naturally integrates predicted watch-time distributions, buffer status, and playlist order, serving as a stable and effective indicator for selecting videos.

\item \textbf{DRL-Enhanced Dynamic Range Selection:} Given the preselected video download order, we design a DRL-based model to determine the appropriate range for each preloading task. The DRL agent is deployed on the client side, enhancing DeLoad's adaptability to dynamic network conditions through reinforcement learning.

\end{itemize}

We train and validate DeLoad through an emulation platform on a dataset collected from real users' sessions with heterogeneous network profiles. Offline evaluations demonstrate that DeLoad achieves 34.4\%–87.4\% improvements in composite QoE scores, reducing rebuffering time by up to 81.4\%. Large-scale online A/B testing over millions of active users confirms these gains, showing 0.9\text{\textperthousand} increase in average watch time per user with bandwidth cost saving by 3.76\% overall. DeLoad has been fully deployed in the commercial short-video platform, Douyin, serving over millions of daily users. To the best of our knowledge, DeLoad is the first production-grade framework to explore dynamic task ranges in short-video preloading with robust watch-time prediction.

%% file: background.tex
\section{Background and Motivation}\label{background}
\begin{figure*}[ht]
    \includegraphics[width=\textwidth]{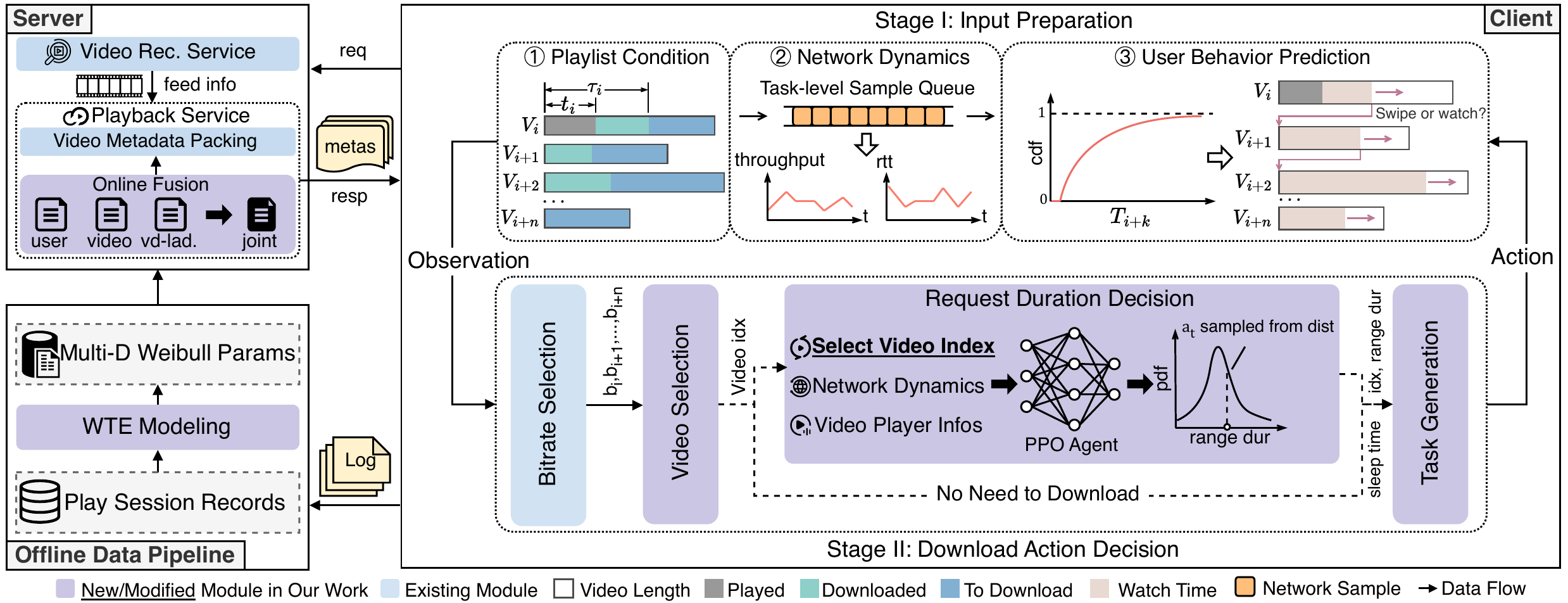}
    \caption{End-to-End System Overview of DeLoad Short-Video Preloading Framework}
    \label{fig_framework}
\end{figure*}


\subsection{Lack of Adaptation to Dynamic Range}
Preload strategy involves at least three key decisions: (1) when to download,  (2) which video and bitrate to download, (3) how much data to download. While many prior works~\cite{chao2022pdas,he2020liveclip,duasvs,dashlet} have focused on the timing and video selection aspects, the critical dimension of HTTP range request granularity determining the optimal data amount per request has remained underexplored. These studies typically assume a fixed-length chunk (in time or size) is fetched after a video is selected for preloading. However, the amount of data to request can significantly influence the effectiveness of any download strategy.
\subsubsection{Range request granularity directly affects throughput, thus impacting playback smoothness}
For long-form videos, studies on HTTP adaptive streaming \cite{dynamic-segment,has-tcp} have shown that throughput is highly sensitive to segment size and request patterns: too many small requests lead to bursty, inefficient bandwidth usage, while larger requests, leading to contiguous data transfer, can better utilize available capacity as shown in the top part of Fig.~\ref{fig:range-request-motivation}. Therefore, when there is a risk of playback stalling, overly small range requests should be avoided to improve bandwidth utilization and ensure smooth playback.
\subsubsection{Range request granularity directly affects scheduling frequency, thus impacting playback smoothness}
Under conventional serialized range request scheduling, subsequent download decisions are triggered only after the completion of previous tasks. As a result, oversized range requests inherently reduce decision-making frequency, leading to slower adaptation to network fluctuations and user behavior changes. As shown in the top part of Fig.~\cite{fig:range-request-motivation}, the number of requests decreases as the range size increases. However, when the range duration exceeds 8 seconds, playback performance begins to degrade significantly, primarily due to the reduced frequency.

\subsubsection{Range request granularity affects bandwidth control, thus impacting bandwidth cost}
Prior studies\cite{measurement-shortvideo,dashlet} have shown that users often swipe quickly through content on short videos, tending to low completion rates. For example, Dashlet\cite{dashlet} reports that in 29\% of video sessions, users swipe within the first 20\% of videos, which is about 3 seconds for short videos. If each video is downloaded in a fixed 5-second range, then for these 29\% of sessions, 2 seconds of data are needlessly fetched, resulting in a substantial waste.

Therefore, to achieve an optimal QoE–bandwidth trade-off, the range request granularity should be dynamically adapted based on factors such as network conditions, user behavior, and bandwidth cost sensitivity. As shown in the bottom part of Fig.~\ref{fig:range-request-motivation}, even under fixed-duration settings, different user cohorts exhibit distinct optimal range lengths (e.g., 4s vs. 7s), driven by varying interaction patterns. Specifically, \textit{casual viewers} refer to users who frequently swipe through content with short watch durations, while \textit{persistent viewers} tend to watch videos to completion with minimal interruptions. This behavioral heterogeneity undermines the effectiveness of \textit{one-size-fits-all} strategies and highlights the need for a context-aware, adaptive scheduling mechanism.

\subsection{Inaccurate and Non-Universal Watch-Time Estimation}

User watch-time is a key factor for making intelligent download decisions. The bottom part of Fig.~\ref{fig:range-request-motivation} illustrates that casual and persistent viewers require significantly different preloading strategies. 
In practice, watch-time is treated as a random variable, and its probability distribution plays a crucial role in determining the download sequence and the expected data wastage. Current approaches have two key limitations.

\subsubsection{Prediction Accuracy.}
Watch-time is influenced by multiple factors, including video content, video length, and user-specific behaviors. Existing works such as PDAS\cite{chao2022pdas} and Dashlet\cite{dashlet} predict watch time or completion rate at the video level, without considering user-level personalization and other context.
\subsubsection{Universality for Large-Scale Platforms.}
Many prior methods\cite{chao2022pdas,dashlet} rely on historical viewing data of each individual video to predict its future watch-time distribution. This approach is ineffective for newly uploaded or less popular videos that lack sufficient viewing history.
As a result, these methods are not broadly applicable in real-world deployments, especially for long-tail content.

%% file: method.tex

\section{Design}\label{sec:method}

\subsection{System Overview}
In this section, we present the technical details of DeLoad, whose end-to-end architecture is illustrated in Fig.~\ref{fig_framework}.

On the server side, DeLoad has two main components. (1)The offline pipeline collects and updates historical session logs. Processed by the Watch Time Estimation (WTE) module detailed in Section~\ref{watch_time_section}, the logs generate multi-dimensional parameters to model watch-time distributions robustly. (2) The online delivery service, composed of video recommendation and playback services, provides video metadata and fused watch-time parameters to clients.

On the client side, DeLoad has two stages. In Stage I, the client constructs a dynamic playlist based on recommended videos. It continuously monitors network conditions, specifically estimating the task-level \textit{throughput} and \textit{RTT} via sliding-window averaging of sampled network statistics. 
Stage II will generate the final preloading decisions. Firstly, bitrate selection (ABR) will decide the playing bitrate which is a dependent module for the whole preloading framework utilizing MPC solution~\cite{mpc2015}. The video selection module (detailed in Section~\ref{video_selection}) determines which video to download next. A configurable buffer threshold parameter $B_{max}$ is applied to prevent excessive bandwidth consumption. 
The range duration decision in Section~\ref{range_model} will decide the task range. Finally, task generation integrates multi-module decisions to generate preloading tasks, automatically triggering pause intervals $T_p$ when preloading becomes counterproductive.

\subsection{Watch-time Distribution Prediction} \label{watch_time_section}

\begin{figure}[t]
    \centering
    \begin{subfigure}[b]{0.23\textwidth}
        \includegraphics[width=\textwidth]{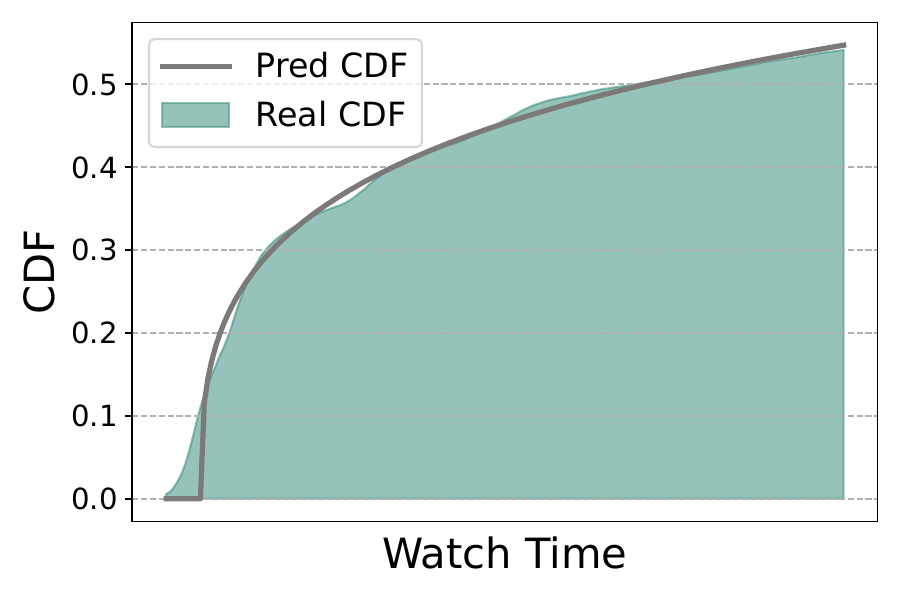}
        \caption{Video Dimension}
        \label{fig_weibull_source}
    \end{subfigure}
    \hfill
    \begin{subfigure}[b]{0.23\textwidth}
        \includegraphics[width=\textwidth]{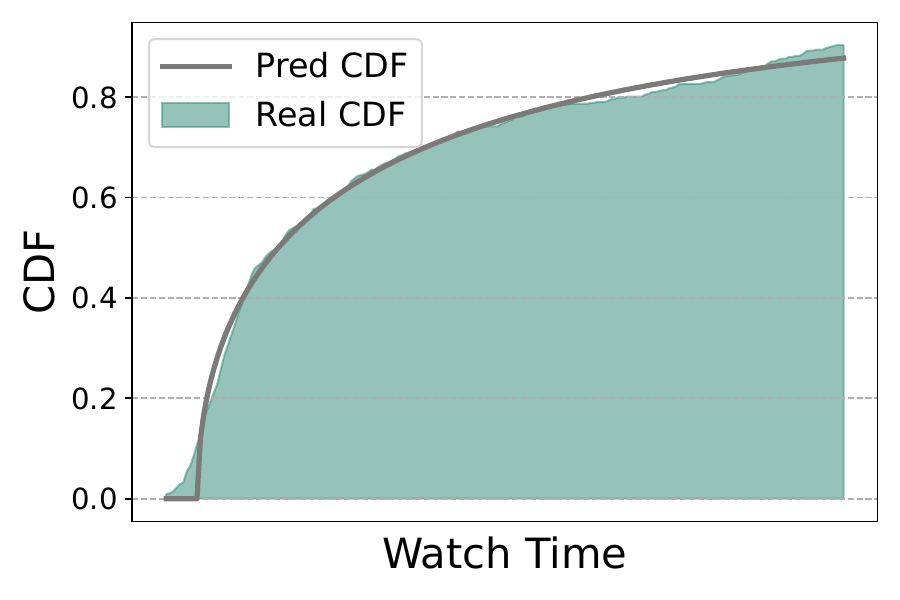}
        \caption{User Dimension}
        \label{fig_weibull_user}
    \end{subfigure}
    
    \vspace{0.5cm}
    
    \begin{subfigure}[b]{0.23\textwidth}
        \includegraphics[width=\textwidth]{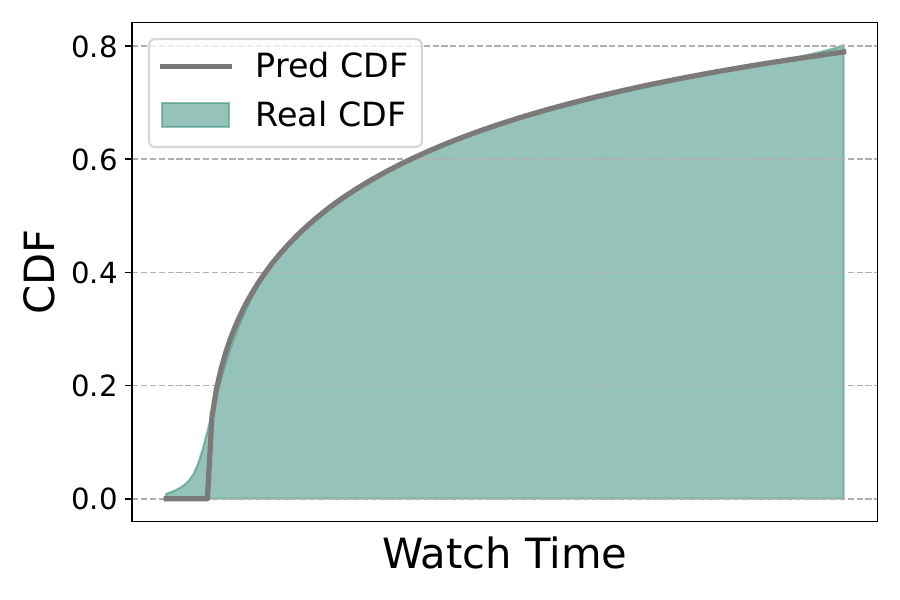}
        \caption{Duration Ladder Dimension}
        \label{fig_weibull_duration}
    \end{subfigure}
    \hfill
    \begin{subfigure}[b]{0.23\textwidth}
        \includegraphics[width=\textwidth]{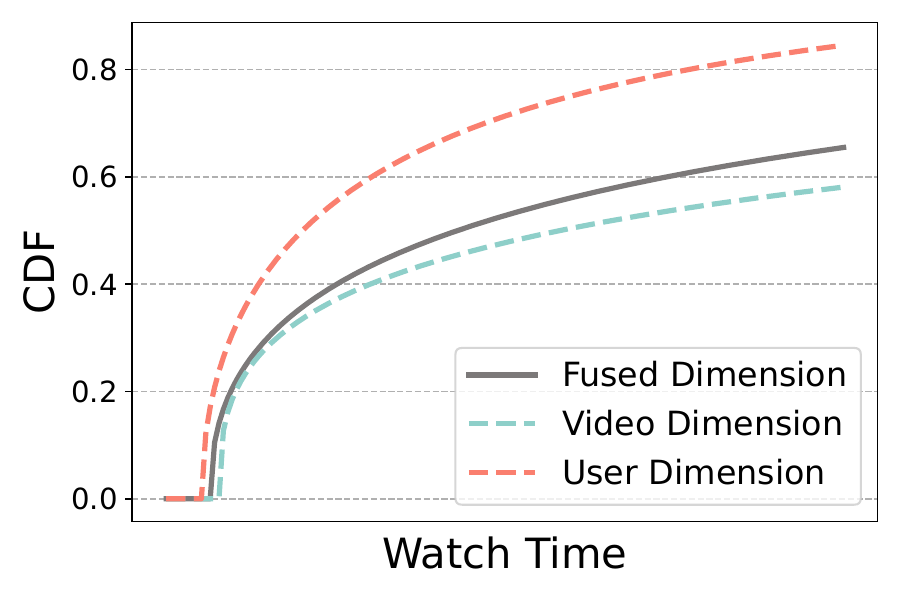}
        \caption{Fused Dimension}
        \label{fig_weibull_final}
    \end{subfigure}
    \caption{Weibull Watch-Time Distribution Overview}
    \label{fig_weibull}
\end{figure}

Accurately predicting a user's watch-time for each video view is crucial for optimizing preloading strategies. When watch-time is known in advance idealy, the optimal strategy is to sequentially download only the segments the user will view, thereby minimizing bandwidth waste while maximizing playback performance (QoE). However, in real-world large-scale platforms, precise prediction of user watch-time remains highly challenging~\cite{watchtime2013}. 

The core difficulties arise from three factors: (\textbf{i}) the massive and heterogeneous user base, (\textbf{ii}) highly diverse and short-lived content, and (\textbf{iii}) non-negligible data latency in production systems and the delay-sensitive nature of streaming pipelines. To address these challenges, we propose an efficient and scalable mechanism for watch-time distribution modeling. This distribution is then integrated into the preloading pipeline through dynamic adaptation, as described in the remainder of this section.

\noindent
\textbf{Why distribution, not scalar?}  
Instead of predicting a single scalar watch-time, we model the full \textit{watch-time distribution} of each video view, which better captures user uncertainty and enables finer-grained preloading strategy. 

\subsubsection{Weibull modeling.}  
We adopt \textit{Weibull analysis} \cite{chao2010dwell}, which is widely used in reliability engineering~\cite{breneman2022introduction}, to model watch-time distribution. Analogous to system failure, a user swiping away from a video can be treated as a “failure” event. Thus, we model this distribution using the Weibull probability density function: 
\begin{equation} \label{eqn-weibull-pdf}
f(t; \beta, \eta, \gamma) = 
\begin{cases}
\frac{\beta}{\eta}\left(\frac{t - \gamma}{\eta}\right)^{\beta - 1} e^{ - \left( \frac{t - \gamma}{\eta} \right)^\beta}, & t > \gamma \\
0, & t \leq \gamma
\end{cases}
\end{equation}

Here, $\gamma$ represents the location parameter, interpreted as the minimum time required for a user to assess the video content and decide whether to continue watching it. $\eta$ is the scale parameter, and $\beta$ controls the shape of the distribution. When $\beta = 1$, the distribution reduces to exponential, indicating a constant swipe probability. A value $\beta < 1$ suggests users are more likely to swipe early, while $\beta > 1$ indicates an increasing likelihood over time.

Besides its interpretability, Weibull modeling also brings \textit{system-level efficiency}: with only three parameters, it minimizes the cost of transmitting and parsing distribution parameters. This stands in contrast to histogram-based methods~\cite{chao2022pdas}, which involve substantial overhead.

\subsubsection{Joint Multi-Dimensional Estimation.}
Due to computational and latency constraints, performing real-time watch-time distribution prediction for every video view is impractical in large-scale systems. To address this, we propose a \textit{Multi-Dimensional Hybrid Framework} that supports offline modeling and online inference, as illustrated in the server component of Fig.~\ref{fig_framework}. The framework consists of two main stages:

\noindent
\textbf{Step 1 — Offline Precomputation.}
We estimate Weibull distribution parameters from three complementary dimensions:

\begin{itemize}[leftmargin=0pt, itemindent=1em]
    \item \textbf{Video Dimension} (Fig.~\ref{fig_weibull_source}): captures how a given video is consumed across the entire user base, reflecting its aggregate attractiveness.
    \item \textbf{User Dimension} (Fig.~\ref{fig_weibull_user}): models personalized watch-time behavior across videos of various durations.
    \item \textbf{Duration Ladder Dimension} (Fig.~\ref{fig_weibull_duration}): serves as a fallback result when user or video data is insufficient, leveraging well-populated reference contents with similar video durations~\cite{watchtime2022kdd}.
\end{itemize}

For each dimension, Weibull parameters are estimated via Least Squares Estimation (LSE)~\cite{ai2000weibullfit}. These models are refreshed periodically to capture recent shifts in user behavior and content trends.

Notably, due to potential data sparsity, LSE fitting for the video and user dimensions may fail. In contrast, the duration ladder dimension consistently has sufficient data coverage, making it a reliable fallback when either of the former two is unavailable.

\noindent
\textbf{Step 2 — Online Fusion.}
At runtime, the final watch-time distribution (Fig.~\ref{fig_weibull_final}) is computed by adaptively fusing the available parameter estimates. If both video and user-specific data are present, we compute the average of the corresponding Weibull parameters. If neither source is missing, such as in the case of new users or long-tail videos, the system falls back to the duration ladder estimate.

The fusion logic is described in Eq.~\ref{eq:weibull-fusion}, where $x$ denotes the Weibull parameters in Eq.~\ref{eqn-weibull-pdf}, i.e., $\beta$, $\eta$, or $\gamma$. The values $x_v$, $x_u$, and $x_d$ represent the estimates from the video, user, and duration ladder dimensions.
\begin{equation}
x_{\text{fused}} =
\begin{cases}
x_d, & x_u = \varnothing \land x_v = \varnothing \\
x_v, & x_u = \varnothing \\
x_u, & x_v = \varnothing \\
\frac{1}{2} \left( x_v + x_u \right), & \text{otherwise}
\end{cases}
\label{eq:weibull-fusion}
\end{equation}

\subsection{Preloading Video Selection} \label{video_selection}

Based on accurate watch-time distribution prediction, we design a download priority metric, denoted as \textbf{Demand}, to guide preloading decisions. By ranking videos in the playlist according to their Demand values, the algorithm can effectively answer the key question in preloading: \textit{which video should be downloaded next}?


\subsubsection{Demand-Based Video Selection}

The preloading playlist typically contains multiple videos, each at a different position in the playback sequence and with varying levels of buffering and predicted watch-time. These factors collectively influence the decision of \textit{which video should be downloaded next}.

We observe that a video whose \textit{uncached portion} is likely to be watched \textit{earlier} should be prioritized for downloading. Based on this insight, we define the concept of \textbf{Demand}, which represents the probability that the uncached portion of video $V_i$ will be the first to be encountered during playback among all videos in playlist.

Since videos are played sequentially, the Demand value for $V_i$ can be expressed as:
\begin{equation}
\text{Demand}(V_i) = P(T_0 \le \tau_0, \dots, T_{i-1} \le \tau_{i-1}, T_i > \tau_i \mid T_0 > t_0)
\label{eqn:demand_general}
\end{equation}

\noindent
where $T_i$ is a \textit{random variable} denoting the watch time for video $V_i$; $\tau_i$ is the currently buffered duration, and $t_0$ is the current playback position for the playing video $V_0$.

\begin{itemize}[leftmargin=0pt, itemindent=1em]
\item \textbf{Playing Video.} For the currently playing video $V_0$, the Demand is simply the probability that the user will watch beyond the current buffer:
\begin{equation}
\text{Demand}(V_0) = P(T_0 > \tau_0 \mid T_0 > t_0)
\label{eqn:demand_playing}
\end{equation}
\item \textbf{Upcoming Videos.} Assuming that user interactions across videos are independent, for any video $V_i$ ($i > 0$), the Demand is defined recursively: it is the probability that all preceding videos are fully buffered for the user's future watching, while $V_i$ is not:

\begin{equation}
\text{Demand}(V_i) = \left(1 - \sum_{k=0}^{i-1} \text{Demand}(V_k) \right) \cdot P(T_i > \tau_i)
\label{eqn:demand_recursive}
\end{equation}
\end{itemize}
This formulation naturally encodes the playback order, watch-time uncertainty, and current buffer state into a unified selection priority metric. In practice, the Demand values are recomputed every time a new download decision is required. The video to download next, denoted as $V_{selected}$, is then determined as:
\begin{align}
V_{selected} = \underset{\substack{\tau_i - t_i < B_{max},~i \in [0, n]}}{\mathrm{argmax}} ~ Demand(V_i)
\label{select_function}
\end{align}
If all buffered durations satisfy $\tau_i - t_i \geq B_{max}$, $V_{selected}$ becomes empty, indicating that sufficient buffer has been achieved for all videos and thus, the downloader can safely enter a sleep state temporarily.

\subsubsection{Demand Calculation}

The core of Demand computation lies in evaluating the probability $P(T_i > \tau_i)$, which measures how likely the user is to watch beyond the currently buffered duration $\tau_i$ for each video $V_i$. As discussed in Section~\ref{watch_time_section}, the watch time $T_i$ is assumed to follow a Weibull distribution, parameterized by $(\beta_i, \eta_i, \gamma_i)$ obtained from the server-side estimation process.
Given this model, the probability that $T_i$ exceeds a certain threshold $\tau_i$ can be computed via the survival function (i.e., $1$ minus the cumulative distribution function, CDF):

\begin{equation}
P(T_i > \tau_i) = 
\begin{cases}
\exp\left( - \left( \frac{\tau_i - \gamma_i}{\eta_i} \right)^{\beta_i} \right), & \tau_i > \gamma_i \\
1, & \tau_i \le \gamma_i
\end{cases}
\label{eqn:survival}
\end{equation}

In the case of the currently playing video $V_0$, the conditional probability in Eq.~\ref{eqn:demand_playing} can be computed as:

\begin{equation}
P(T_0 > \tau_0 \mid T_0 > t_0) = \frac{P(T_0 > \tau_0)}{P(T_0 > t_0)}
\label{eqn:conditional}
\end{equation}

Combining Eq.~\ref{eqn:survival} and Eq.~\ref{eqn:conditional}, we can compute the complete set of Demand values across all videos in the playlist. These probabilities are updated at every decision point using the most recent playback position and buffer status.

\subsection{Dynamic Video Range Duration Model} \label{range_model}

\begin{figure}[t]
    \centering
    \includegraphics[width=0.485\textwidth]{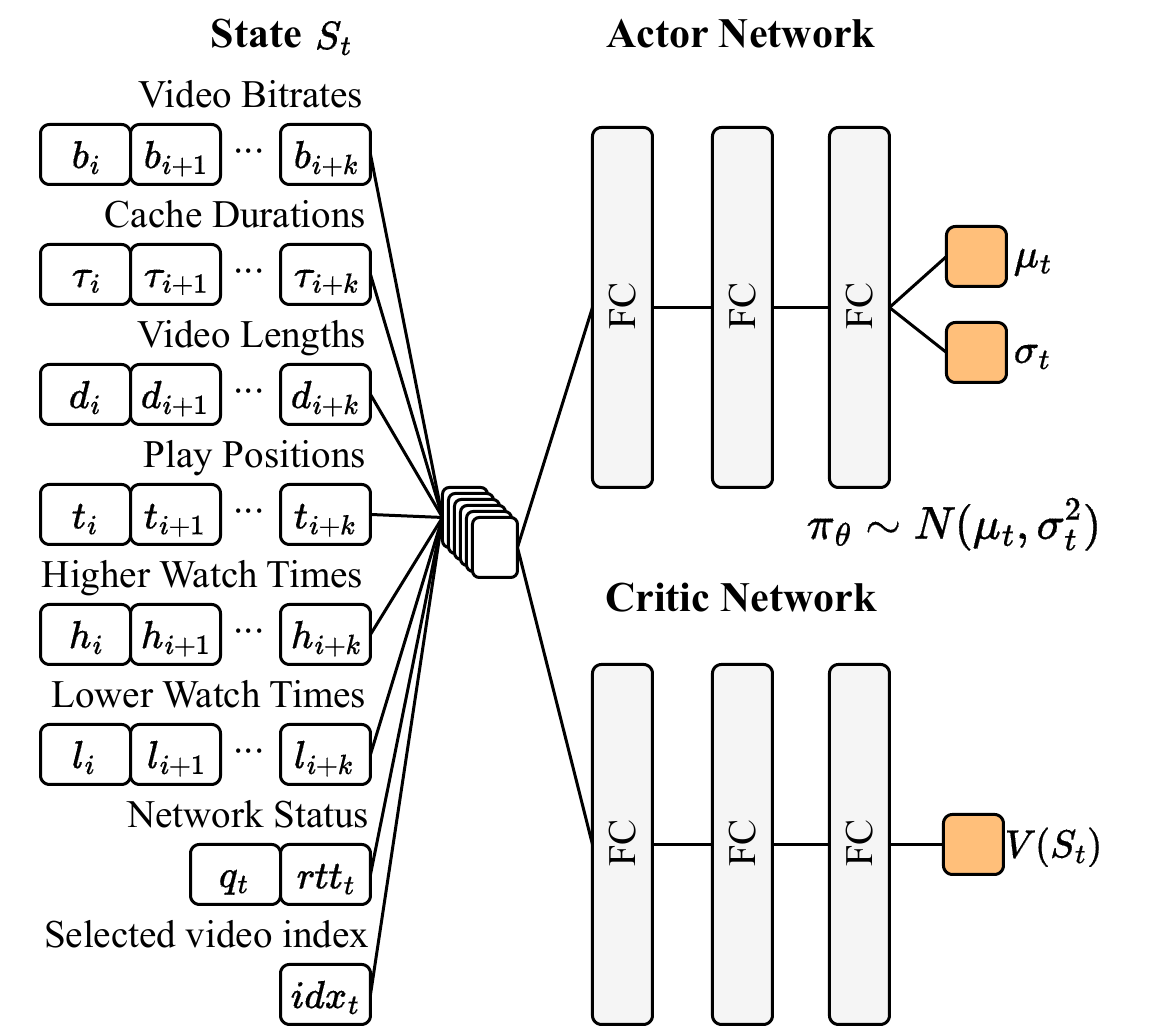}
    \caption{Architecture of Dynamic Video Range Model}
    \label{fig_model}
\end{figure}

In this section, we present \textbf{Dynamic Video Range Model}, which determines the range action results upon receiving the video's Demand values. As elaborated in Section~\ref{background}, fixed range downloading proves inadequate in balancing the QoE and minimizing data waste. This inadequacy motivates our exploration of transitioning from finite discrete actions to continuous actions. However, developing a heuristic algorithm within a continuous action space to address this problem, considering the vast input states from the playlist, is extremely challenging. Consequently, we employ a DRL framework to create a more effective range agent. Specifically, DeLoad is trained with PPO~\cite{ppo} algorithm, which is one of the most valuable Reinforcement Learning frameworks and has demonstrated excellent performance in numerous studies~\cite{ppoiot,ppoabr,gpt,zhang2024bbq,fan2025aether,fan2025echocc}.

\subsubsection{Input state} The states of our model take into account the finite videos in the playlist. This generated task will be assigned to $V_{selected}$, which has the highest Demand value and satisfies the buffer limit $B_max$. Additionally, the following videos' information is also vital for this decision-making. For the playlist with $n$ videos, we choose the top $k$ videos to construct the input, where $k \leq n$. For each video $V_i$, we form the state tuple as $[b_i, \tau_i, d_i, t_i, h_i, l_i]$, and this tuple is normalized separately for each video. Here, $b_m$ is the video's bitrate, $\tau_i$ denotes the video's cached duration, and $d_m$ is the original video's length, which is a fixed value. $t_i$ is the play position of this video. $h_i, l_i$ are two distinct estimated play time results which are generated by two hyper-parameters $e_h$ and $e_l$, representing the higher and lower ratio of estimated watch time, both within $[0, 1.0)$. Separating the estimated play time results into two input states enables the collection of more play time information from Weibull watch-time distribution and incorporation into the decision-making model to compensate inefficiencies of the original watch time prediction. Moreover, we integrate the network dynamic information into the model, including throughput $q_t$ and Round Trip Time $rtt_t$. The last state is the index $idx_t$ of $V_{selected}$, whose index starts from the current playing video.

\subsubsection{Agent's action} Range model's action $a_t$ is a continuous output representing the range duration. Actor network's outputs are the mean value and standard deviation of Normal Distribution, defined as $\mu_t$ and $\sigma_t$. We generate the action output distribution as $\pi_{\theta} \sim \mathcal{N}(\mu_t, \sigma_t^2)$. In practical deployment, we will apply this range decision to the video $V_{selected}$. Hence, this action mainly represents the buffer data urgency for the selected video.

\subsubsection{Reward function} Reward function of range model consists of three components: bitrates, data waste, and rebuffering. First of all, we want to maximize the bitrate $b_t$ results for better user experience and reduce the cached data waste $w_t$ for each range action $a_t$. $w_t$ is the unwatched video size in the buffer after the user swipes away. Additionally, rebuffering time $bt_t$ is another crucial factor affecting QoE, which times throughput $q_t$ to represent the needed data. Besides, introducing throughput $q_t$ can balance the different reward values under higher and lower network environments. Thus, we combine these three key factors into one reward function to trade off with different hyper-parameters. We define $r$ as follows:
\begin{align}
r & = a_t * b_t - \alpha * w_t - \beta * bt_t * q_t.
\label{range_reward_func}
\end{align}
During model training, we define a trajectory as the completion of the entire playlist on a single network trace. The accumulated reward is $R = \sum_t{r}$. The loss function of the actor network we used is $L^{CLIP}(\theta)$~\cite{ppo}. The architecture details of the dynamic video range model are shown in Fig.~\ref{fig_model}.

\subsection{Implementation on Mobile Device}\label{imple}
We present different modules' settings and implementation details in this section. 
On the server side, we update the data pipeline records each day for the watch-time distribution. On the client-side, the dynamic video range model in DeLoad takes $k=5$ videos' information for states. For the tail videos in the playlist, we add the default zero values for alignment. And the higher and lower watch time ratio is $e_h=0.7$ and $e_l=0.3$. The parameters in the reward function are $\alpha = 0.01, \beta = 1.85$. And the waste data $w_t$ is clipped by the maximal value 1.2M in bits empirically. The actor network adopts fully connected layers. Specifically, input states pass two hidden layers with 128 and 64 neurons with a ReLU function. Then the output layer has two neurons, of which the mean value $\mu$ uses tanh function and $\sigma$ uses softplus function. The critic network has a similar network structure but has one neuron for the output layer without an activation function, which estimates the state value. Learning rate is  $1*10^{-6}$ for both networks. 
The model was trained and optimized using the PyTorch framework under experimental conditions. For deployment, the trained model was converted into the standardized ONNX format to ensure cross-framework compatibility and integrated with a dedicated machine learning SDK for client-side execution across heterogeneous hardware platforms (CPU/GPU/NPU). The lightweight machine learning SDK, implemented in C++, facilitates on-device machine learning task execution through a Python runtime environment supported by a lightweight virtual machine to handle complex algorithmic logic. This hybrid architecture enables computationally intensive operations to leverage native C++ efficiency while maintaining Python's flexibility for algorithm customization. Additional algorithmic components, video selection, and task generation are deployed as native C++ modules within the client's player in the application, with inter-module communication achieved through standardized APIs. This modular design ensures efficient data flow between feature processing, model inference, and post-processing stages while maintaining hardware abstraction. The task generation module will pause the downloader for $T_p=500ms$ before the next generation if all preloading videos are found exceeding $B_{max} = 10$s.

%% file: evaluation.tex
\section{Evaluation}\label{evaluation}

\begin{figure*}[t]
    \centering
    \begin{subfigure}[b]{0.32\textwidth}
        \includegraphics[width=\textwidth]{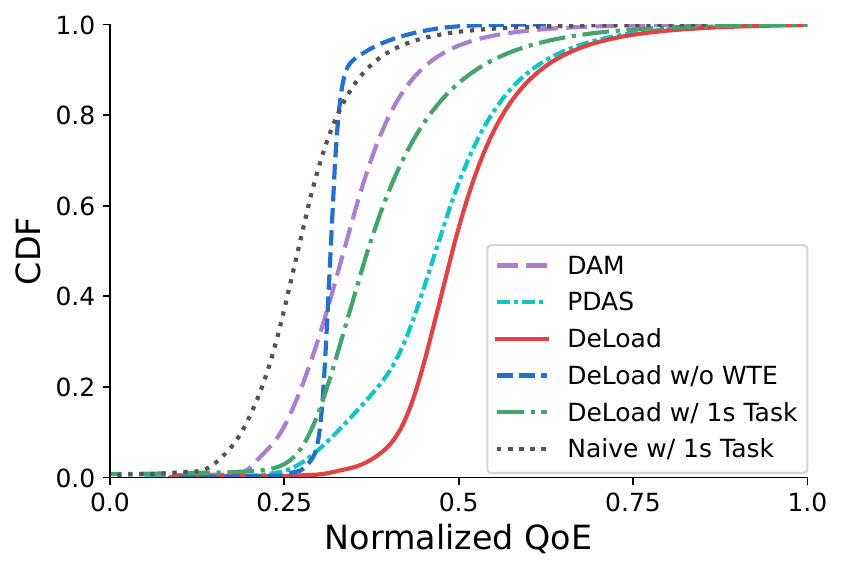}
        \caption{Evaluation Overall Normalized QoE CDF}
        \label{offline_cdf}
    \end{subfigure}
    \hfill
    \begin{subfigure}[b]{0.32\textwidth}
        \includegraphics[width=\textwidth]{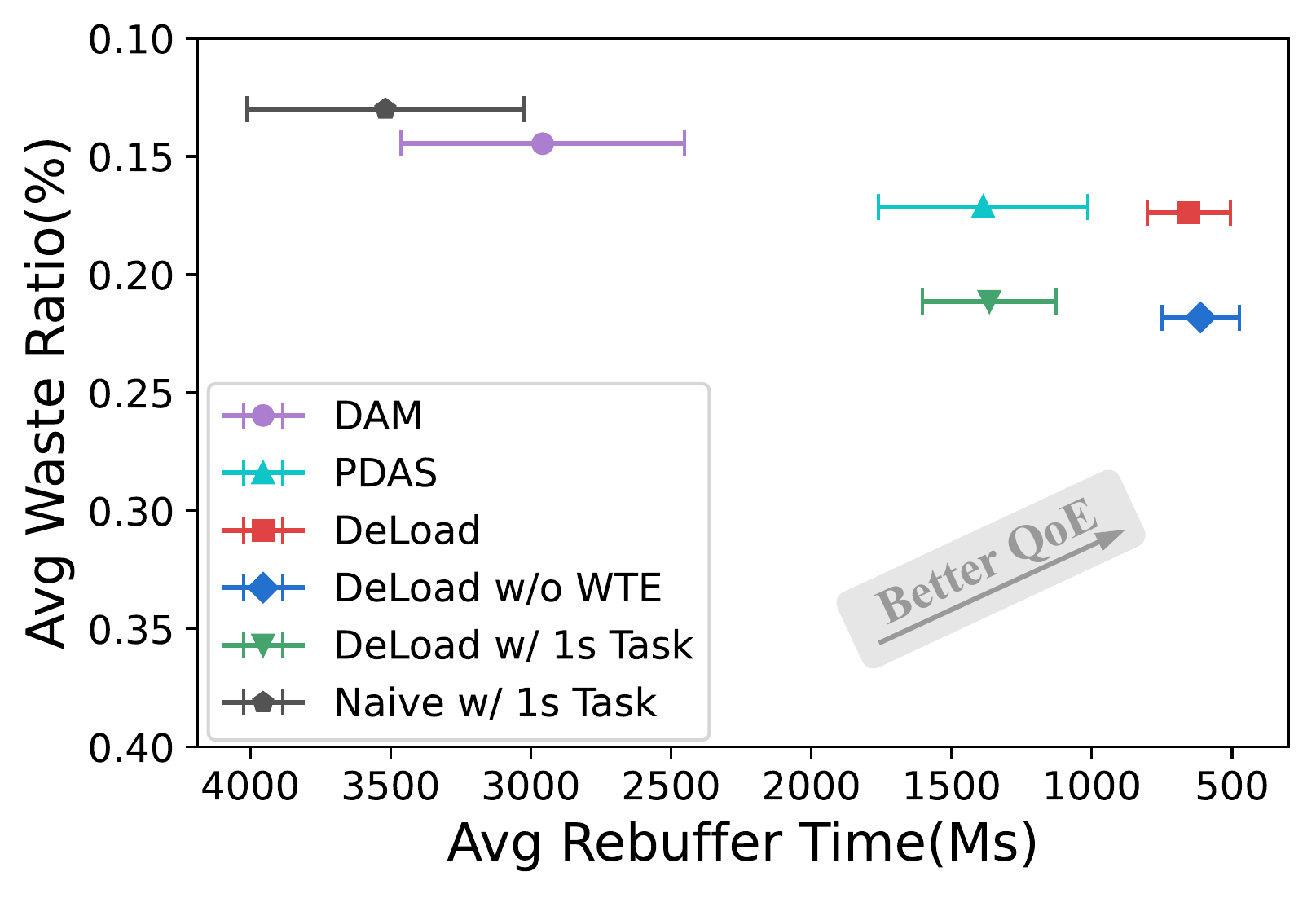}
        \caption{Rebuffering vs Waste Results}
        \label{offline_qos}
    \end{subfigure}
    \hfill
    \begin{subfigure}[b]{0.32\textwidth}
        \includegraphics[width=\textwidth]{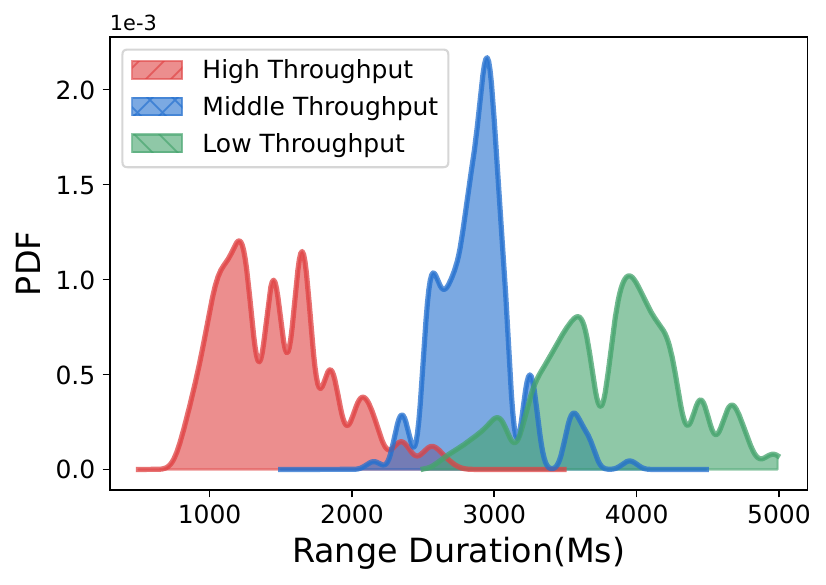}
        \caption{Range Duration Distribution of DeLoad}
        \label{offline_action}
    \end{subfigure}
    \caption{Offline Evaluation Results}
    \label{fig:offline_evaluation}
\end{figure*}

In this section, we comprehensively evaluate the performance of DeLoad both offline and online. For the offline evaluation, we utilize the modified testing platform for training and evaluation. In the online phase, we conduct A/B tests on Douyin over several weeks. 

\subsection{Offline Methodology}

\subsubsection{Baseline Algorithms.}\label{algorithms_section}
\begin{itemize}[leftmargin=0pt, itemindent=1em]

    \item \textbf{PDAS}~\cite{chao2022pdas}: PDAS is a probability-driven adaptive downloading algorithm to optimize the bandwidth waste and QoE simultaneously. It consumes the watch time probability to describe the swiping events in second granularity. Based on this, it determines the preloading order, bitrates, and sleeping time results. 
    \item \textbf{DAM}~\cite{qian2022dam}: It is a short-video preloading algorithm based on domain knowledge and DRL, which shares a similar PPO framework. It decides the download video, target bitrate level, and pause flag. It uses a fixed request duration for each task, such as 1s.
    \item \textbf{DeLoad w/o WTE}: It is a simplified DeLoad version without \textbf{W}atch-\textbf{T}ime \textbf{E}stimation for evaluating. It includes two main modifications: use the average watch-time data to calculate a uniform distribution for all videos to calculate the Demand value in the video selection and remove the higher and lower watch-time estimation values from states in the range model.
    \item \textbf{DeLoad w/ 1s Task}: It is another simplified DeLoad version that still uses the video selection algorithm to determine the one video with the highest Demand value but downloads a fixed 1s task.
    \item \textbf{Naive w/ 1s Task}: A heuristic preloading algorithm, each time preloading a fixed 1s task for the first video with insufficient buffer.
\end{itemize}

\subsubsection{Simulation Setup} During the offline evaluation, we leverage online data to train and define our models. Specifically, we gather nearly 3,000 network traces and simulate over 150,000 actual hours of playing multiple online videos. We use an open testing platform to evaluate different algorithms. To better adapt to the online environment, we add two rational modifications based on the open version. On the one hand, to satisfy dynamic range preloading tasks, we modify it from a fixed 1s step simulation to a flexible step length, such as 100ms. On the other hand, we change the fixed RTT for each download task to a random RTT value. This random RTT is sampled from a Continuous Uniform Distribution $U(40, 120)$, ranging from 40ms to 120ms. In this simulation platform, there are 5 videos in the default recommendation queue for each playlist. It reads the user's retention rate from the dataset to simulate the user's swiping action. 
Besides, this testbed is a trace-driven simulation process. It plays back all the videos until the last video swiping away on each network trace. As our training reward function~\ref{range_reward_func} design, we use it as our offline QoE metrics. The hyper-parameters are the same as the implementation in Section~\ref{imple}.

\subsection{Offline Evaluation}

\begin{figure*}[t]
    \centering
    \begin{subfigure}[b]{0.49\textwidth}
        \includegraphics[width=\textwidth]{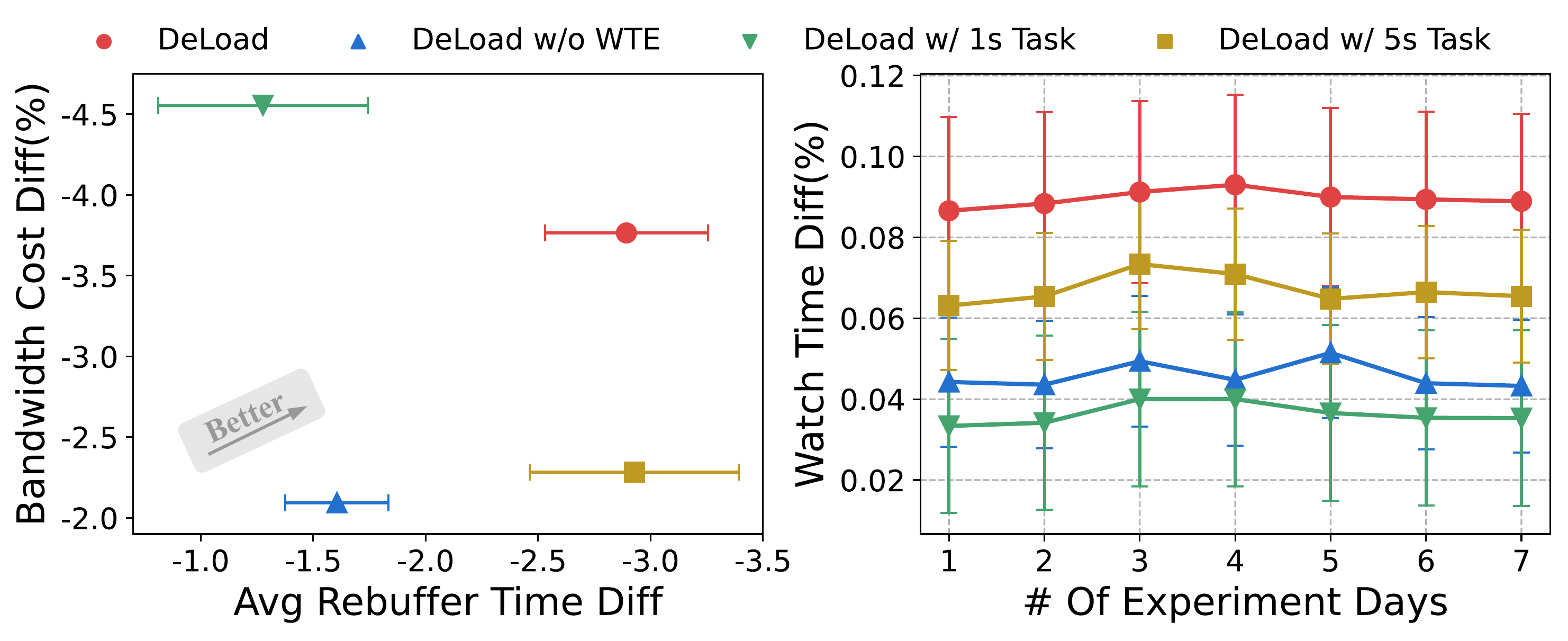}
        \caption{Overall Performance and User Watch-Time Diff in One Week}
        \label{fig:online_overall}
    \end{subfigure}
    \hfill
    \label{fig:online_evaluation}
    \begin{subfigure}[b]{0.49\textwidth}
        \includegraphics[width=\textwidth]{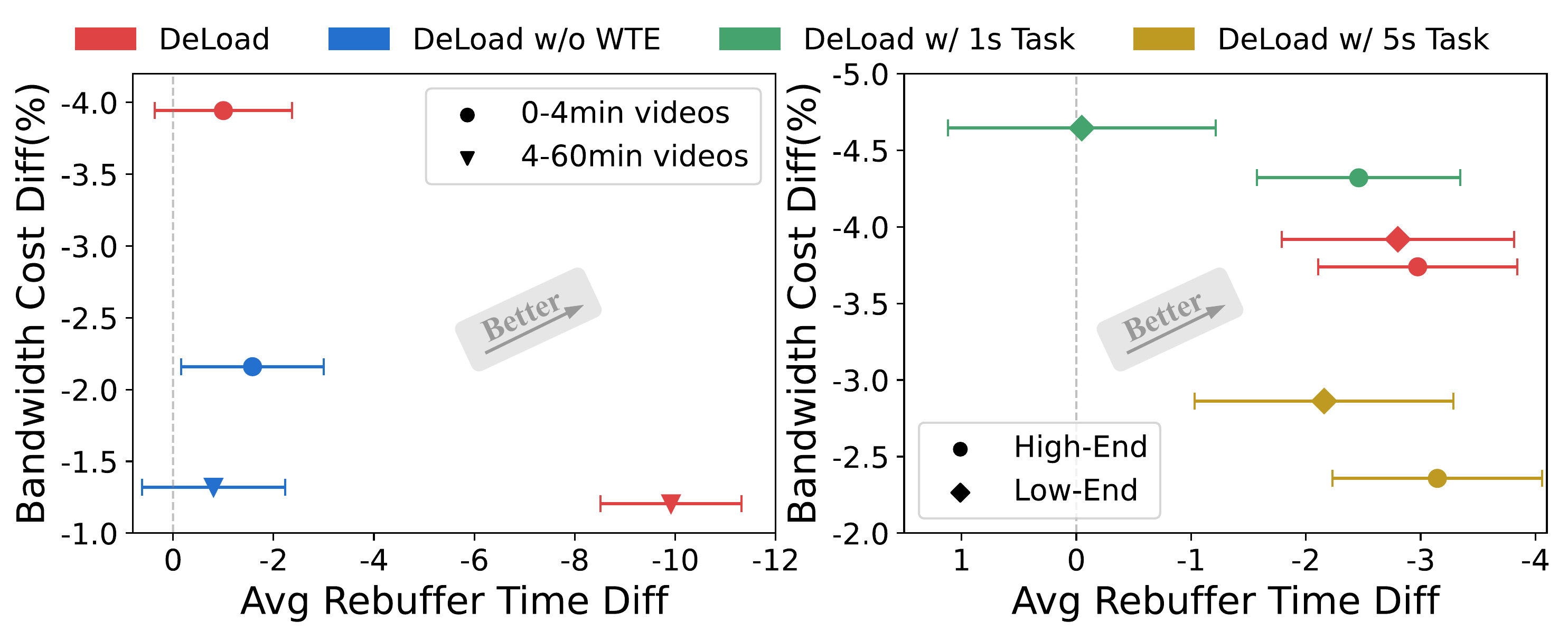}
        \caption{Performance Across Video Lengths and Device Capabilities}
        \label{fig:online_detail}
    \end{subfigure}
    \caption{Online Evaluation Results}
\end{figure*}

\subsubsection{Overall Performance Results}
Fig.~\ref{offline_cdf} shows the overall QoE metrics results. We can see that DeLoad exhibits the best results among all the algorithms, with improvements of 34.4\%-87.4\%. In contrast, Naive w/ 1s Task performs worst. PDAS outperforms DAM. This is because PDAS models the users' actions more reasonably. In comparison, DAM merely incorporates the watch probability into its state input. Furthermore, compared with DeLoad w/ 1s Task and DeLoad, the dynamic range model achieves a significant improvement with 64.3\% of QoE. It benefits from the DRL model with online data to tackle these dynamic decision tasks. Specifically, DeLoad w/o WTE shows that the absence of the watch-time distribution mechanism leads to a relatively lower QoE, highlighting that the full DeLoad's watch-time distribution mechanism contributes to a 76.4\% improvement in QoE.

\subsubsection{Rebuffering vs. Waste Results Analysis}
Fig.~\ref{offline_qos} illustrates the rebuffering and data waste results. From two-dimensional plots, DeLoad shows significant improvements in rebuffering results with 52.1\% - 81.4\%. Interestingly, DeLoad w/o WTE's rebuffering time is less than the completed DeLoad accompanied by a higher waste ratio. This is because the model will generate a larger range to ensure QoS for the playlist without the watch-time distribution, which is an inefficient bandwidth allocation strategy. Watch-time estimation can enable the algorithm to estimate the future playing process to balance the two metrics. DeLoad shows nearly the top performance in data waste compared with other algorithms, while Naive w/ 1s Task gets the lowest data waste. As the analysis in our motivation, small preloading sizes are better for data waste and worse for rebuffering. Thus, Naive w/ 1s Task gets the worst rebuffering results but the best data waste performance. DAM and PDAS show similar results with Naive w/ 1s Task but a better trade-off between rebuffering and data waste. PDAS achieves a greater optimization than the two above algorithms, but worse than DeLoad. Specifically, DeLoad achieves about 34.4\% rebuffering time improvement with almost the same waste ratio result as PDAS. For a commercial platform, such substantial QoS improvement under the same level of bandwidth cost can accomplish a higher business revenue, which shown in Section~\ref{online_evaluation}. The ablation version of DeLoad, DeLoad w/ 1s Task, shows worse in both rebuffering and data waste. This is because flexible ranges can reduce the network connection time compared to several series 1s tasks, which is particularly advantageous with low-buffer levels.

\subsubsection{Understanding Dynamic Range Results}
To gain a more in-depth understanding of the dynamic range mechanism, we collect the action results during the operation of DeLoad and estimate the Probability Density Function(PDF) in Fig.~\ref{offline_action}. It is evident that DeLoad adapts its range decisions according to diverse network conditions. When the user has a high throughput, the task size is typically smaller for better bandwidth efficiency. On the contrary, when the throughput is low, the user requires a larger range to ensure smooth video playback, even if there is a high risk of creating more data waste. In the majority of cases, as shown by the middle throughput, the range model will generate a more suitable range size for preloading. This trade-off may be dynamic along with the network's conditions. Overall, our range model demonstrates remarkable robustness in handling this task, effectively optimizing the preloading range to adapt to various network scenarios.

\subsection{Online A/B Tests}\label{online_evaluation}

We implemented and deployed four variants—\textbf{DeLoad w/ 1s Task}, \textbf{DeLoad w/ 5s Task}, \textbf{DeLoad w/o WTE}, and \textbf{DeLoad} as described in Section~\ref{algorithms_section}. We then conducted a comprehensive A/B experiment to compare their performance against the platform’s original preloading strategy, which serves as the baseline.

Fig.~\ref{fig:online_overall} summarizes the results over 7 consecutive days, showing how the four methods perform relative to the baseline in terms of rebuffering time, bandwidth cost, and, ultimately, their impact on overall user watch-time, which is one of the key business metrics for short video platforms.

We observe that the full version approach DeLoad achieves the best overall watch-time (improving by \textbf{0.9\text{\textperthousand}} over the baseline), with near-optimal playback smoothness and the most competitive bandwidth cost, demonstrating a favorable trade-off between playback quality and bandwidth efficiency.
\subsubsection{Study on Adaptive Range Request Granularity}
We compare DeLoad w/ 1s Task, DeLoad w/ 5s Task, and DeLoad to examine the impact of request granularity in Fig.~\ref{fig:online_overall}.
\begin{itemize}[leftmargin=0pt, itemindent=1em]
\item DeLoad w/ 1s Task performs the best in bandwidth cost due to its small download size (1 second). Still, it significantly underutilizes available bandwidth, resulting in the highest rebuffering time and ultimately the worst user watch time.

\item DeLoad w/ 5s Task achieves a playback smoothness similar to DeLoad, but at the cost of higher bandwidth waste.

\item Notably, DeLoad matches DeLoad w/ 5s Task in rebuffer time but excels in total watch time.
\end{itemize}
To further investigate the performance difference across user segments, we stratified users into \textit{high-end} and \textit{low-end} device groups based on hardware specifications. While \textit{DeLoad w/ 5s Task} achieves slightly lower average rebuffering time across all users compared to \textit{DeLoad}, a more detailed analysis reveals that \textbf{DeLoad provides much better playback smoothness for low-end device users} as shown in the right part of Fig.~\ref{fig:online_detail}.

Low-end users typically suffer from both weaker hardware and poorer network conditions. The improved performance of \textit{DeLoad} in this group likely stems from its ability to dynamically assign larger download tasks when the risk of stalling is high, thereby utilizing available bandwidth more effectively in challenging environments. We believe this adaptive behavior is a key factor contributing to its superior performance in overall watch time.
These results demonstrate the practical value of dynamically adapting range request size in real-world short video streaming systems.

\subsubsection{Study on the Role of Watch-Time Prediction}
We also compare DeLoad w/o WTE with DeLoad to isolate the impact of accurate watch-time distribution prediction.
\begin{itemize}[leftmargin=0pt, itemindent=1em]
\item Across all metrics, DeLoad outperforms DeLoad w/o WTE, benefiting from a more accurate and robust estimation of watch-time distribution.

\item A breakdown by video length, as shown in the left part of Fig.~\ref{fig:online_detail}, that the smoothness improvement of DeLoad is most apparent for longer videos, while its bandwidth-saving advantage is more significant for shorter videos. This further confirms the value of precise watch-time prediction and its effective integration into preloading strategies for short video applications.
\end{itemize}

%% file: relatedwork.tex
\section{Related Work}\label{related_work}

\noindent\textbf{Preloading.} To optimize user experience in short video applications and reduce bandwidth costs, many efforts have been invested in preloading strategies. 
Current methods can be broadly classified into two main paradigms: rule-based approaches (e.g., heuristic methods in Dashlet\cite{dashlet}, Lyapunov optimization in APL\cite{zhang2020apl}, and MPC in PADS\cite{chao2022pdas}), and learning-based approaches that utilize deep (reinforcement) learning to autonomously derive preloading strategies, as demonstrated by DAM\cite{qian2022dam}, Alfie\cite{alfie}, DUASVS\cite{duasvs}, Quty\cite{quty}, and others.
These methods primarily focus on optimizing the granularity of fixed-size video chunks, with an emphasis on determining which subsequent chunk to preload.

\noindent\textbf{User Swipe Prediction.} A key challenge in preloading is the uncertainty of user swipes. 
To mitigate the impact of random swiping, recent advancements have attempted to explicitly model user viewing duration, as demonstrated by methods like PDAS\cite{chao2022pdas}, Dashlet\cite{dashlet}, GAMORA\cite{gamora}, DTAAP\cite{Digitaltwin}, PaCs\cite{PACS}, and Joint\cite{joint}. 
These approaches leverage platform-collected video-level viewing duration data to predict swipe behavior and guide preloading decisions.
However, in this work, we highlight significant limitations in video-level viewing duration prediction and propose a user-level viewing duration prediction method.

%% file: conclusion.tex
\section{Conclusion}\label{sec:conclusion}
In this paper, we propose a novel demand-driven framework, DeLoad, for short-video preloading. This framework comprises three main modules: watch-time estimation, preloading video selection, and DRL-enhanced dynamic range model. We conduct comprehensive evaluations of DeLoad and deploy it in a practical framework. Our extensive offline and online experimental results unequivocally demonstrate that DeLoad outperforms existing approaches in terms of performance metrics. Moreover, it has been proven to generate substantial business revenue for the platform.

\begin{acks}
We appreciate the kind instructions of Xiaocheng Li and Deliang Fu during this project. And we extend our gratitude to our colleagues during engineering development - Tianze Yu, Qing Huang, Xiaoyi Xu and Ruigang Wang.
This work is supported in part by the National Science Foundation of China under No. 62225204 and 62472041, in part by the A3 Foresight Program of NSFC under Grant 62061146002.
\end{acks}